\documentclass[12pt]{article}
\usepackage{amsmath,amssymb}
\usepackage{epsfig,graphicx}

\def\APJ{ Astroph.~J.~}
\def\AJ{ Astron.~J.~}
\newcommand{\Fref}[1]{Figure~\ref{#1}}

\newcommand{\arccosh}{\mathop{\mathrm{arccosh}}}

\begin{document}

\title{\textbf{Construction of Exact Solutions in Two-Fields Models and the Crossing
of the Cosmological Constant Barrier}}
\author{S.~Yu.~Vernov\footnote{{\bf e-mail}:
svernov@theory.sinp.msu.ru}\\
\em Skobeltsyn Institute of Nuclear Physics, Moscow State
University\\ {\em Vorobyevy Gory, Moscow, 119991, Russia}}
\date{}
\maketitle

\begin{abstract}
A dark energy model with a phantom scalar field, an usual scalar
field and the string field theory inspired polynomial potential
has been constructed. A two-parameter set of exact solutions to
the Friedmann equations has been found. We have construc\-ted such
stringy inspired potential that some exact solutions correspond to
the state parameter $w_{DE}^{\vphantom{27}}>-1$ at large time,
whereas other ones correspond to $w_{DE}^{\vphantom{27}}<-1$ at
large time. We demonstrate that the superpotential method is very
effective to seek new exact solutions. We also present a
two-fields model with a polynomial potential and the state
parameter, which crosses the cosmological barrier infinitely
often.
\end{abstract}

\section{Introduction}
One of the most important recent results of the observational
cosmology is the conclusion that the Universe is speeding up
rather than slowing down. The combined analysis of the type Ia
supernovae, galaxy clusters measurements and WMAP (Wilkinson
Microwave Anisotropy Probe) data gives strong evidence for the
accelerated cosmic expansion~\cite{Riess1,Spergel}.

The cosmological acceleration suggests that the present day
Universe is dominated by a smoothly distributed slowly varying
cosmic fluid with negative pressure, the so-called dark energy
(DE)~\cite{Tegmark,Astier,Copeland,Gong,ASS,VaStar}. To specify a
component of a cosmic fluid one usually uses a phenomenological
relation between the pressure $p$ and the energy density $\varrho$
corresponding to each component of fluid
 $p=w\varrho$. The function $w$ is called as the state parameter.
 Contemporary
experiments~\cite{Riess1,Spergel,Tegmark,Astier} give strong
support that the Universe is approximately spatially flat and the
DE state parameter $w_{DE}^{\vphantom {27}}$ is currently close to
$-1$:
 \begin{equation}
w_{DE}^{\vphantom {27}}={}-1\pm 0.2.
\end{equation}

The state parameter $w_{DE}^{\vphantom {27}}\equiv -1$ corresponds
to the cosmological constant. From the theoretical point of view
(see~\cite{VaStar,AKV} and references therein) this domain of
$w_{DE}^{\vphantom {27}}$ covers three essentially different
cases: $w_{DE}^{\vphantom {27}}>-1$, $w_{DE}^{\vphantom
{27}}\equiv {}-1$ and $w_{DE}^{\vphantom {27}}<-1$. From the
observations there is no barrier between these three
possibilities. Moreover it has been shown in~\cite{VarunStar} that
the state parameter $w_{DE}^{\vphantom {27}}$, which gives the
best fit to the experimental data, evolves from $w_{DE}^{\vphantom
{27}}\simeq 0$ to $w_{DE}^{\vphantom {27}}\leqslant -1$ and for a
large region in parameter space an evolving state parameter
$w_{DE}^{\vphantom {27}}$ is favoured over $w_{DE}^{\vphantom
{27}}\equiv -1$.

The standard way to obtain an evolving state parameter
$w_{DE}^{\vphantom {27}}$ is to include scalar fields into a
cosmological model.  Under general assumptions within single
scalar field four-dimensional models one can realize only one of
the following possibilities: $w_{DE}^{\vphantom {27}}\geqslant-1$
(quintessence models) or $w_{DE}^{\vphantom {27}}\leqslant-1$
(phantom models)~\cite{Vikman}. Two-fields models with the
crossing of the cosmo\-logical constant barrier $w_{DE}^{\vphantom
{27}}\equiv{}-1$ are known as quintom models and include one
phantom scalar field and one usual scalar field. Note that the
most of phenomenological models describing the crossing of the
cosmological constant barrier~\cite{across-1,Andrianov,AK} use a
few scalar fields or a modified gravity.

Nowadays string and D-brane theories have found cosmological
applications related to the acceleration of the Universe. In
phenomenological models, describing the case $w_{DE} < -1$, all
standard energy conditions are violated and there are problems
with stability at classical and quantum levels
(see~\cite{Copeland,wless-1,AreVo} and references therein).
Possible way to evade the instability problem for models with
$w_{DE}<-1$ is to yield a phantom model as an effective one, which
arises from more fundamental theory with a normal sign of a
kinetic term. In particular, if we consider a model with higher
derivatives such as $\phi e^{-\Box}\phi$, then in the first
nontrivial approximation we obtain $\phi
e^{-\Box}\phi\simeq\phi^2-\phi\Box\phi$, and such a model gives a
kinetic term with a ghost sign. It turns out, that such a
possibility does appear in the string field theory (SFT)
framework~\cite{Arefeva} (see also~\cite{AK,AreVo}), namely in the
theory of fermionic NSR string with GSO$-$ sector. According to
Sen's conjecture (see~\cite{SFT-review} for review), the scalar
field $\phi$ is an open string theory tachyon, describing the
brane decay. The four dimensional gravitational model with a
phantom scalar field is considered as a string theory
approximation, that gives a possibility to solve instability
problems.

In this paper we consider a SFT inspired gravitational model with
two scalar fields and a polynomial potential, which is a
generalization of a one-field cosmological model, describing
in~\cite{AKV}. The first two-fields generalization of this
one-field model has been proposed in~\cite{AKV3} as a polynomial
model, which has a one-parameter set of exact solutions with the
state parameter $w_{DE}$, which crosses the barrier
$w_{DE}^{\vphantom {27}}=-1$ at large time and reaches $-1$ from
below at infinity. In this paper we construct a new model with a
two-parameter set of exact solutions, for some values of
parameters we obtain $w_{DE}^{\vphantom {27}}<-1$ at large time,
whereas for other $w_{DE}^{\vphantom {27}}>-1$ at large time. Note
that the different  behavior of $w_{DE}^{\vphantom {27}}$ at large
time corresponds to one and the same potential and asymptotic
conditions of the fields.

We study different possibility to use the superpotential method
and demonstrate that it is very useful not only to construct
potential for the given exact solutions, but also to seek new
exact solutions. To demonstrate that the superpotential method
allows to find a form of a polynomial potential and solutions for
the given Hubble parameter we construct a toy two-fields model for
the Hubble parameter proposed in the SFT inspired model with high
derivatives~\cite{AK}.

\section{String Field Theory Inspired Two-Fields Model}

We consider a model of Einstein gravity interacting with a single
phantom scalar field $\phi$ and one standard scalar field $\xi$ in
the spatially flat Friedmann Universe. In typical cases a phantom
scalar field represents the open string tachyon, whereas the usual
scalar field corresponds to the closed string
tachyon~\cite{Arefeva,AKV3,AJ,LY}. Since the origin of the scalar
fields is connected with the string field theory the action
contains the typical string mass $M_s$ and a dimensionless open
string coupling constant $g_o$:
\begin{equation}
S=\!\int\! d^4x \sqrt{-g}\left(\frac{M_P^2}{2M_s^2}R+
\frac1{g_o^2}\left(\frac1{2}g^{\mu\nu}(\partial_{\mu}\phi\partial_{\nu}\phi
-\partial_{\mu}\xi\partial_{\nu}\xi) -V(\phi,\xi)\right)\right),
\label{action}
\end{equation}
where $M_P$ is the Planck mass. The Friedmann metric $g_{\mu\nu}$
is a spatially flat:
\begin{equation*}
ds^2={}-dt^2+a^2(t)\left(dx_1^2+dx_2^2+dx_3^2\right),
\end{equation*}
where $a(t)$ is a scale factor. The coordinates $(t,x_i)$ and
fields $\phi$ and $\xi$ are dimensionless.

 If the scalar fields depend only on time,
then the equations of motion are as follows
\begin{eqnarray}
&H^2=\displaystyle\frac{1}{3m_p^2}\left({}-\frac12\dot\phi^2
+\frac12\dot\xi^2+V\right),
\label{eom1}
\\
&\dot
H=\displaystyle\frac{1}{2m_p^2}\left(\dot\phi^2-\dot\xi^2\right),
\label{eom2}
\\
&\ddot\phi+3H\dot\phi=\displaystyle{}\frac{\partial
V}{\partial\phi}, \label{eom3}
\\
&\ddot\xi+3H\dot\xi=\displaystyle{}-\frac{\partial
V}{\partial\xi}. \label{eom4}
\end{eqnarray}
For short hereafter we use the dimensionless parameter $m_p$:
$m_p^2=g_o^2M_P^2/M_s^2$. Dot denotes the time derivative. The
Hubble parameter $H\equiv \dot a(t)/a(t)$. Note that only three of
four differential equations (\ref{eom1})--(\ref{eom4}) are
independent. Equation (\ref{eom4}) is a consequence of
(\ref{eom1})--(\ref{eom3}).

The DE state parameter can be expressed in terms of the Hubble
parameter:
\begin{equation}
\label{w} w_{DE}=-1-\frac23\frac{\dot H}{H^2}.
\end{equation}
The crossing of the cosmological constant barrier $w_{DE}=-1$
corresponds to change of sign of $\dot H$. The phantom like
behavior corresponds to an increasing Hubble parameter.  If we
know the explicit form of fields $\phi(t)$ and $\xi(t)$ and do not
know the potential $V(\phi,\xi)$, then, using eq.~(\ref{eom2}), we
can obtain $H(t)$ with an accuracy to a constant:
\begin{equation}
\label{Ht2}
    H(t)=\frac{1}{2m_p^2}\left(\int\limits^t\dot\phi^2(\tau) d\tau
    -\int\limits^t\dot\xi^2(\tau) d\tau\right) +C.
\end{equation}

At the same time if we know $H(t)$ we can find the potential as a
function of time:
\begin{equation}
\label{Vt}
    V(t)=m_p^2\left(3H(t)^2+\dot H(t)\right).
\end{equation}

The Aref'eva DE model~\cite{Arefeva} (see
also~\cite{AKV,AK,AKV3,AKV2,AKVBulg}) assumes that our Universe is
a slowly decaying D3-brane and its dynamics is described by the
open string tachyon mode. To describe the open string tachyon
dynamics a level truncated open string field theory is used. The
notable feature of such tachyon dynamics is a non-local polynomial
interaction~\cite{SFT-review,ABKM1,ABKM2,Witten-SFT,AMZ-PTY,BSZ}.
It has been found that the open string tachyon behavior is
effectively modelled by a scalar field with a negative kinetic
term~\cite{AJK}. In this paper we consider local models with
effective potentials $V(\phi,\xi)$. The form of these potentials
are assumed to be given from the string field theory within the
level truncation scheme. Usually for a finite order truncation the
potential is a polynomial and its particular form depends on the
string type. The level truncated cubic open string field theory
fixes the form of the interaction of local fields to be a cubic
polynomial with non-local form-factors. Integrating out low lying
auxiliary fields one gets the fourth degree
polynomial~\cite{ABKM1,ABKM2}. Higher order auxiliary fields may
change the coefficients of lower degree terms and produce higher
degree monomials.

The back reaction of this brane is incorporated in the dynamics of
the closed string tachyon. The scalar field $\xi$ comes from the
closed string sector, similar to~\cite{Oh} and its effective local
description is given by an ordinary kinetic term~\cite{LY} and,
generally speaking, a non-polynomial self-interaction~\cite{BZ}.
An exact form of the open-closed tachyon interaction is not known
and, following~\cite{AKV3}, we consider the simplest polynomial
interaction.

More exactly we impose the following restrictions on the potential
$V(\phi,\xi)$:
\begin{itemize}
\item the potential is the sixth degree polynomial:
\begin{equation}
\label{potenV}
 V(\phi,\xi)=\sum_{k=0}^{6}\sum_{j=0}^{6-k}c_{kj}\phi^k\xi^j,
\end{equation}

\item coefficient in front of the fifth and sixth powers are of
order $1/m_p^2$ and the limit $m_p^2\to \infty$ gives a nontrivial
fourth degree potential,

\item the potential is even: $V(\phi,\xi)=V(-\phi,-\xi)$. It means
that if $k+j$ is odd, then $c_{kj}=0$.

\end{itemize}

From the SFT we can also assume asymptotic conditions for
solutions. To specify the asymptotic conditions for scalar fields
let us recall that we have in mind the following picture. We
assume that the phantom field $\phi(t)$ smoothly rolls from the
unstable perturbative vacuum ($\phi=0$) to a nonperturbative one,
for example $\phi=1$, and stops there. The field $\xi(t)$
corresponds to close string and is expected to go asymptotically
to zero in the infinite future. In other words we seek such a
function $\phi(t)$ that $\phi(0)=0$ and it has a non-zero
asymptotic at $t\to +\infty$: $\phi(+\infty)=A$. The function
$\xi(t)$ should have zero asymptotic at $t\to +\infty$. At the
same time we can not calculate the explicit form of solutions in
the string field theory framework.

In this paper we show how using the superpotential method we can
construct a potential and exact solutions, which satisfy
conditions obtaining in the SFT framework.

\section{The Method of Superpotential}

The gravitational models with one or a few scalar fields play an
important role in cosmology and theories with extra dimensions.
One of the main problems in the investigation of such models is to
construct exact solutions for the equations of motion. System
(\ref{eom1})--(\ref{eom4}) with a polynomial potential
$V(\phi,\xi)$ is not integrable. Moreover we can not integrate
even models with one scalar field and a polynomial potential.

The superpotential method has been proposed for construction of a
potential, which corresponds to the exact solutions to
five-dimensional gravitational models~\cite{DeWolfe}. The main
ideas of this method are to consider the function H(t) (the Hubble
parameter in cosmology) as a function (superpotential) of scalar
fields and to construct the potential for the special solutions,
given in the explicit form.
 Let
\begin{equation}
H(t)=W\bigl(\phi(t),\xi(t)\bigr).
\end{equation}
Equation (\ref{eom2}) can be rewritten as follows
\begin{equation}
\frac{\partial W}{\partial\phi}\dot\phi+\frac{\partial
W}{\partial\xi}\dot\xi=\frac1{2m_p^2}\left(\dot\phi^2-\dot\xi^2\right).
\label{alexey_W_equation}
\end{equation}

If one find such $W(\phi,\xi)$ that the relations
\begin{equation}
\dot\phi=2m_p^2\frac{\partial W}{\partial\phi},
\label{deWolfe_method1}
\end{equation}
\begin{equation}
 \dot\xi={}-2m_p^2\frac{\partial W}{\partial\xi},
\label{deWolfe_method2}
\end{equation}
\begin{equation}
V =3m_p^2W^2+2m_p^4\left(\left(\frac{\partial W}{\partial
\phi}\right)^2-\left(\frac{\partial W}{\partial
\xi}\right)^2\right) \label{deWolfe_potential}
\end{equation}
are satisfied, then the corresponding $\phi(t)$, $\xi(t)$ and
$H(t)$ are a solution of system (\ref{eom1})--(\ref{eom4}).

The superpotential method separates system
(\ref{eom1})--(\ref{eom4}) into two parts: system
(\ref{deWolfe_method1})--(\ref{deWolfe_method2}), which is as a
rule integrable for the given polynomial $W(\phi,\xi)$ and
equation (\ref{deWolfe_potential}), which is not integrable if
$V(\phi,\xi)$ is a polynomial, but has a special polynomial
solutions. The way to use of superpotential method does not
include solving of eq.~(\ref{deWolfe_potential}). The potential
$V(\phi,\xi)$ is constructed by means of the given $W(\phi,\xi)$.

There are a few ways to use the superpotential method. The
standard way~\cite{DeWolfe} is to construct the potential for the
solutions given in the explicit form. We assume an explicit form
of solutions, find the superpotential $W$ and use
(\ref{deWolfe_potential}) to obtain the corresponding potential
$V$. If we consider one-field models, putting, for example,
$\xi\equiv 0$ in (\ref{eom1})--(\ref{eom4}), then from
(\ref{deWolfe_method1}) we obtain $W(\phi)$ up to a constant. At
the same time solving (\ref{deWolfe_method1}) we obtain a
one-parameter set of solutions: $\phi(t-t_0)$. So in the case of
one-field models we have the following correspondence
\begin{equation}
      \phi(t-t_0) \quad \leftrightarrow \quad W(\phi)+C,
      \label{phiW}
\end{equation}
where $t_0$ and $C$ are arbitrary constants.

In two-fields models the correspondence (\ref{phiW}) does not
exist and the superpotential method gives a possibility to find
new solutions. Indeed,
equations~(\ref{deWolfe_method1})--(\ref{deWolfe_method2}) form
the second order system of differential equations. If this system
is integrable then we obtain two-parameter set of solutions. To
assume some explicit form of solutions means to assign a
one-parameter set of solutions. The superpotential method allows
to generalize this set of solutions up to two-parameter set.  On
the other hand we can construct different forms of superpotential
and potential, which correspond to one and the same one-parameter
set of solutions.

The idea to consider the Hubble parameter as a function of scalar
fields and to transform (\ref{eom1})--(\ref{eom4}) into
(\ref{deWolfe_method1})--(\ref{deWolfe_potential}) has been used
in the Hamilton--Jacobi formulation of the Friedmann
equations~\cite{Muslimov,Salopek} (see also~\cite{Liddle}) and
does not connect with supersymmetric and supergravity theories. At
the same time the method, based on the idea to apply system
(\ref{deWolfe_method1})--(\ref{deWolfe_potential}) instead of the
original equations of motion for the search exact special
solutions, is actively used in two-dimensional fields
models~\cite{Bazeia95,Bazeia99} and
supergravity~\cite{Brandhuber}.
Equations~(\ref{deWolfe_method1})--(\ref{deWolfe_method2}) are
known as the Bogomol'nyi equations~\cite{Bogomolnyi} (see
also~\cite{Bazeia99}). The superpotential method is a combination
and a natural extension of these two methods. This method is
actively used in cosmology~\cite{AKV,AKV3,BazeiaDE,BazeiaCDM}. Let
us note generalizations of this method on the equations of motion,
describing the close and open Friedmann universes~\cite{BazeiaDE},
systems with the cold dark matter~\cite{BazeiaCDM} and the
Brans--Dicke theory~\cite{MMVS}. The idea to consider the Hubble
parameter as a function of scalar fields and to transform
(\ref{eom1})--(\ref{eom4}) into
(\ref{deWolfe_method1})--(\ref{deWolfe_potential}) has been used
in the Hamilton--Jacobi formulation of the Friedmann
equations~\cite{Muslimov,Salopek} (see also~\cite{Liddle}) and
does not connect with supersymmetric and supergravity theories. At
the same time the idea to apply system
(\ref{deWolfe_method1})--(\ref{deWolfe_potential}) instead of the
original equations of motion and to seek in such a way exact
special solutions is actively used in two-dimensional fields
models~\cite{Bazeia95,Bazeia99}, supergravity~\cite{Brandhuber}
and supersymmetric models with the BPS states.
Equations~(\ref{deWolfe_method1}) and (\ref{deWolfe_method2}) are
known as the Bogomol'nyi equations~\cite{Bogomolnyi} (see
also~\cite{Bazeia99}). The superpotential method is a combination
and a natural extension of these two methods. At present the
superpotential method is actively used in
cosmology~\cite{AKV,AKV3,BazeiaDE,BazeiaCDM}. Let us note
generalizations of this method on the equations of motion for the
close and open Friedmann universes~\cite{BazeiaDE}, systems with
the cold dark matter~\cite{BazeiaCDM} and the Brans--Dicke
theory~\cite{MMVS}.

\section{The construction of potentials for the given solutions}

\subsection{Non-polynomial potential}
In this section we demonstrate that one and the same solutions can
correspond to both polynomial and non-polynomial potentials. In
the next section we show that the superpotential method allows to
find different exact solutions, which correspond to the different
behavior of the Hubble parameter, but one and the same potential.

From the asymptotic conditions we assume the following explicit
form of solution:
\begin{equation}
\label{sol2} \phi(t)=A\tanh(\omega t) \qquad \mbox{and}\qquad
\xi(t)=\frac{A\sqrt{2(1+b)}}{\cosh(\omega t)},
\end{equation}
where $A>0$, $\omega>0$ and $b>-1$.

From (\ref{Ht2}) we obtain
\begin{equation}
\label{Ht1}
 H(t)=\frac{A^2\omega }{6m_p^2}\Bigl(3\tanh(\omega t)-(3+2b)\tanh^3(\omega t)\Bigr).
\end{equation}

Note that this kink-lump solution is a natural generalization of
the kink solution for the one-field phantom model~\cite{AKV}.  The
behavior of the Hubble parameter at large time depends on the
parameter $b$. From the contemporary experimental data it follows
that the present date Universe is expanding one that corresponds
to $H>0$ at large time. The condition $\lim\limits_{t=+\infty}H>0$
is equivalent to $b<0$. Eventually, we state that $-1<b<0$.

On the other hand, in the past there were eras of the accelerated
and decelerated expanding Universe, it means that the Hubble
parameter $H$ has to be not a monotonic function and should has an
extremum at some point $t_c>0$. From (\ref{Ht1}) we obtain that
\begin{equation}
t_{c}=\frac{1}{\omega}\arccosh\left(\pm
\frac{\sqrt{2(b+1)(2b+3)}}{2(b+1)} \right). \label{tc0}
\end{equation}
We have assumed that $b>-1$, so the sign "+" corresponds to real
$t_{c}$. At $t>0$ the Hubble parameter $H$ has one extremum,
namely a maximum. The corresponding DE state parameter $w_{DE}$ is
given by
\begin{equation}
w_{DE}=-1+12m_p^2\frac{\left(2(b+1)\cosh(\omega
t)^2-3-2b\right)\cosh(\omega t)^2} {A^2\sinh(\omega
t)^2(2b\sinh(\omega t)^2-3)^2}.
\end{equation}
It is easy to see that at large time $w_{DE}>-1$, so we obtain the
quintessence like behavior of the Universe\footnote{It has been
shown~\cite{AKVBulg} that if we consider the other pair of scalar
fields
\begin{equation}
\label{sol2bulg} \tilde{\phi}(t)=\tanh(t), \qquad
\tilde{\xi}(t)=\frac{\sqrt{(1+b)}}{\cosh(2 t)},
\end{equation}
then for some values of parameter $b$, for example $b=-0.01$, the
corresponding Hubble parameter has both a maximum and a minimum at
$t>0$ and increases at large time. Note that the polynomial
potential, which corresponds to solutions~(\ref{sol2bulg}), is not
known.}.

Let us construct a potential, which corresponds to fields
(\ref{sol2}). The functions $\phi(t)$ and $\xi(t)$ are solutions
of the following system of differential equations:
\begin{equation}
\label{equphixi} \left\{
\begin{split}
    \dot\phi&=A\omega-\frac{\omega}{A}\phi^2, \\
    \dot\xi&=\omega\xi\sqrt{1-\frac{\xi^2}{2(1+b)A^2}}.
\end{split}
\right.
\end{equation}

The straightforward use of the superpotential method gives
\begin{equation}
\label{DW} \frac{\partial
W}{\partial\phi}=\frac{\omega}{2m_p^2}\Bigl(A-\frac{1}{A}\phi^2\Bigr),
\qquad \frac{\partial
W}{\partial\xi}={}-\frac{\omega\xi}{2m_p^2}\sqrt{1-\frac{\xi^2}{2(1+b)A^2}}.
\end{equation}

Therefore,
\begin{equation}
\label{W1}
   H\equiv W=\frac{\omega}{6m_p^2}\left(3A\phi-\frac{\phi^3}{A}-
    \sqrt{\frac{\left(2(1+b)A^2-\xi^2\right)^3}{2(1+b)A^2}} + H_0\right),
\end{equation}
where $H_0$ is an arbitrary constant. Different values of $H_0$
correspond to different $V(\phi,\xi)$. The obtained potentials
\begin{equation}
\label{V1}
 V=\omega^2\left(\frac{\left(A^2-\phi^2\right)^2}{2A}
-\frac{\xi^2}{2}+ \frac{\xi^4}{4(1+b)A^2}\right)+3m_p^2W^2
\end{equation}
are polynomial ones only in the flat space-time ($m_p^2=\infty$)
and do not satisfy conditions of Section 2.

The goal of this paper is to construct a polynomial potential
model with such set of exact solutions that the quintessence large
time behavior  corresponds to some solutions and the phantom large
time behavior  corresponds to other ones. The potential and
solutions should satisfy conditions from Section 2. In other words
our model should be the SFT inspired one. We make this
construction in two steps. At the first step we construct
polynomial potential for (\ref{sol2}). At the second step we find
new solutions for the obtained polynomial potential.

\subsection{New polynomial potentials for the given solutions} Let us
construct for the functions (\ref{sol2}) such a superpotential
that the corresponding potential has the polynomial form.
Functions (\ref{sol2}) satisfy not only system (\ref{equphixi}),
but also the following system of differential equations:
\begin{equation}
\left\{
\begin{split}
\dot\phi&=\displaystyle A\omega b\left(\frac{\phi^2}{A^2}-1\right)+\frac{\omega\xi^2}{2A},\\
 \dot\xi&=\displaystyle {}-\frac{\omega}{A}\phi\,\xi.
\end{split}
\right.
 \label{time_dependence-anzats-1}
\end{equation}

The corresponding Hubble parameter (superpotential) is  given by
\begin{equation}
H=\tilde{W}=\frac{\omega\phi}{2m_p^2}\left(Ab\left(\frac{\phi^2}{3A^2}-1\right)+
\frac{\xi^2}{2A}\right)+H_0. \label{W2}
\end{equation}

To obtain even potential we put $H_0=0$:
\begin{equation}
\tilde{V}=\frac{\omega^2}{2}\left(b\left(\phi^2-1\right)+\frac{1}
{2}\xi^2\right)^2 -\frac{\omega^2}{2A^2}\phi^2\xi^2+
\frac{3\omega^2\phi^2}{4m_p^2}\left(Ab\left(\frac{\phi^2}{3A^2}-1\right)+
\frac{\xi^2}{2A}\right)^2. \label{alexey_V}
\end{equation}

This example shows that the same functions $\phi(t)$ and $\xi(t)$
can correspond to essentially different potentials $V(\phi,\xi)$.
So, we conclude that in two-fields models one has more freedom to
choose the potential, without changing solutions than in one-field
models. Moreover, the solutions do not change if we add to the
potential $\tilde{V}$ (or $V$) a function $\delta V$, which is
such that $\delta V$, $\partial(\delta V)/\partial\phi$ and
$\partial(\delta V)/\partial\xi$ are zero on the solution. For
example, we can add
\begin{equation}
\delta
V=K(\phi,\xi)\left[\phi^2+\frac{1}{2(1+b)}\xi^2-A^2\right]^2,
\label{deltaV}
\end{equation}
where $K(\xi,\phi)$ is a smooth function. So, we can obtain new
potentials, which correspond to the given exact
solutions~(\ref{sol2}), without constructing of new
superpotentials.

\section{Construction of new solutions via the superpotential method}

In previous section we have shown how we can choose potential for
the given solutions. In this section we demonstrate the
possibility to find new exact solutions (may be in quadratures)
using superpotential method. Let us consider the model with the
potential (\ref{alexey_V}). It is easy to see that system
(\ref{time_dependence-anzats-1}) has not only solutions
(\ref{sol2}), but also the trivial solutions $\{\phi(t)=\pm A,
\quad \xi(t)=0\}$ and solution
\begin{equation}
\label{solxi0}
    \phi(t)=-A\tanh(\omega b(t-t_0)), \qquad \xi(t)=0.
\end{equation}

If $\xi(t)\not\equiv 0$, then, using the second equation of
(\ref{time_dependence-anzats-1}), we obtain the second order
differential equation in $\xi(t)$:
\begin{equation}
\label{equ2xi}
   \ddot \xi(t)=\omega^2b\,\xi(t)-\frac{\omega^2 \xi^3(t)}{2A^2}
   +\frac{(1-b)\dot \xi^2(t)}{\xi(t)}.
\end{equation}

The solutions of eq. (\ref{equ2xi}) with $b>-1$ are defined in
quadratures
\begin{equation}
\label{solxi}
  t-t_0=\pm\int\frac{A\sqrt{2(1+b)}\xi^{b-1}}
  {\omega\sqrt{2A^2\xi^{2b}+2A^2b\xi^{2b}+\xi^{2b+2}+2A^2C+2A^2bC}}\,d\xi,
 \end{equation}
where $C$ and $t_0$ are arbitrary  constants. For some values of
parameter $b$ the general solution to
(\ref{time_dependence-anzats-1}) can be written in the explicit
form, for example at $b={}-1/2$ we obtain:
\begin{equation}
\label{solphixi_1_2}
\begin{split}
    \phi(t)&=\frac{A\left(\left(C_1^2C_2^2+4A^2\right)e^{\omega t}
    -C_1^2e^{-\omega t}\right)}
    {\left(C_1^2C_2^2+4A^2\right)e^{\omega t}+2C_1^2C_2+C_1^2e^{-\omega t}},\\
\xi(t)&= \frac{4C_1A^2}{\left(C_1^2C_2^2+4A^2\right)e^{\omega
t}+2C_1^2C_2+C_1^2e^{-{\omega t}}},
\end{split}
\end{equation}
where $C_1$ and $C_2$ are arbitrary parameters. It is easy to
check that for all values of $C_1$, but $C_1=0$, and $C_2$
solutions (\ref{solphixi_1_2}) and the Hubble parameter $H(t)$
satisfy the following asymptotic conditions:
\begin{equation}
    \phi(\pm\infty)=\pm A,\qquad \xi(\pm\infty)=0,\qquad
    H(\pm \infty)={}\pm \frac{A^2\omega}{6m_p^2}.
\end{equation}

So we have constructed a gravitational model with a two-parameter
set of exact solutions. The potential and solutions satisfy
conditions, imposed by means of the string field theory (see
Section 2).

Let us analyze the property of the obtained solutions and the
cosmological consequences. System (\ref{time_dependence-anzats-1})
is invariant to change $\xi(t)$ on $-\xi(t)$, so each solution
$\phi(t)$ corresponds to two solutions ${}\pm\xi(t)$. Note that
the function $\phi(t)$ is invariant to the change $C_1 \rightarrow
{}-C_1$, whereas the function $\xi(t)$ changes a sign. The Hubble
parameter depends on $\xi^2$, so, without loss of generality, we
can put $C_1>0$.

System~(\ref{time_dependence-anzats-1}) is autonomous one, so if
there exists a solution $\{\tilde{\phi}(t), \tilde{\xi}(t)\}$,
then a pair of functions
$\{\tilde{\phi}(t-t_0),\tilde{\xi}(t-t_0)\}$, where $t_0\in
\mathbb{C}$, has to be a solution as well. It is convenient to use
in (\ref{solphixi_1_2}) such parameters that one of them
corresponds to a shift of solutions in time. We put
$C_1=exp(t_0)$. Using the restriction $t_0\in \mathbb{R}$ we come
to the condition $C_1>0$. For short we introduce the new parameter
$C\equiv C_1C_2$ instead of $C_2$. Solutions  (\ref{solphixi_1_2})
take the following form:
\begin{equation} \label{phixi+C}
\begin{split}
    \phi(t)&={}\frac{A\left((C^2+4A^2)e^{\omega(t-t_0)}-e^{-\omega(t-t_0)}\right)}
    {(C^2+4A^2)e^{\omega(t-t_0)}+2C+e^{-\omega(t-t_0)}},\\ \xi(t)&=
\frac{{}4A^2}{(C^2+4A^2)e^{\omega(t-t_0)}+2C+e^{-\omega(t-t_0)}}.
\end{split}
\end{equation}

To compare the obtained solutions with the initial solution
(\ref{sol2}), we introduce new parameter  $t_1\equiv t_0+t_{00}$,
where
\begin{equation}
\label{t00}
t_{00}\equiv{}-\frac{1}{2\omega}\ln\left(C^2+4A^2\right).
\end{equation}

 Now functions $\phi(t)$ and $\xi(t)$ are
\begin{equation}
\label{phixi2}
\begin{split}
    \phi(t)&={}\frac{A\left(e^{\omega(t-t_1)}-e^{-\omega(t-t_1)}\right)}
    {e^{\omega(t-t_1)}+\frac{2C}{\sqrt{C^2+4A^2}}+e^{-\omega(t-t_1)}},\\
    \xi(t)&=
\frac{4A^2}{\sqrt{C^2+4A^2}\left(e^{\omega(t-t_1)}+\frac{2C}{\sqrt{C^2+4A^2}}
+e^{-\omega(t-t_1)}\right)}.
\end{split}
\end{equation}
Let us consider solutions with $t_1=0$. It is easy to see that in
this case
\begin{equation}
   \phi(0)=0,\qquad \dot\phi(0)=\frac{A\omega\sqrt{C^2+4A^2}}{C+\sqrt{C^2+4A^2}}>0
    \qquad \mbox{and} \qquad \dot\xi(0)=0.
\end{equation}
From (\ref{eom2}) it follows that $\dot H(0)>0$ and from
(\ref{W2}) it follows that $H(0)=0$. Therefore, solutions with
$t_1=0$ and an arbitrary $C$ are cosmological bounce solution (see, for
example,~\cite{Cai07}), in other words, $a(t)$ has a bounce in the point $t=0$.

Let us consider how the behavior of the Hubble parameter $H(t)$
depends on $C$.

\begin{figure}[h]
\centering
\includegraphics[width=50mm]{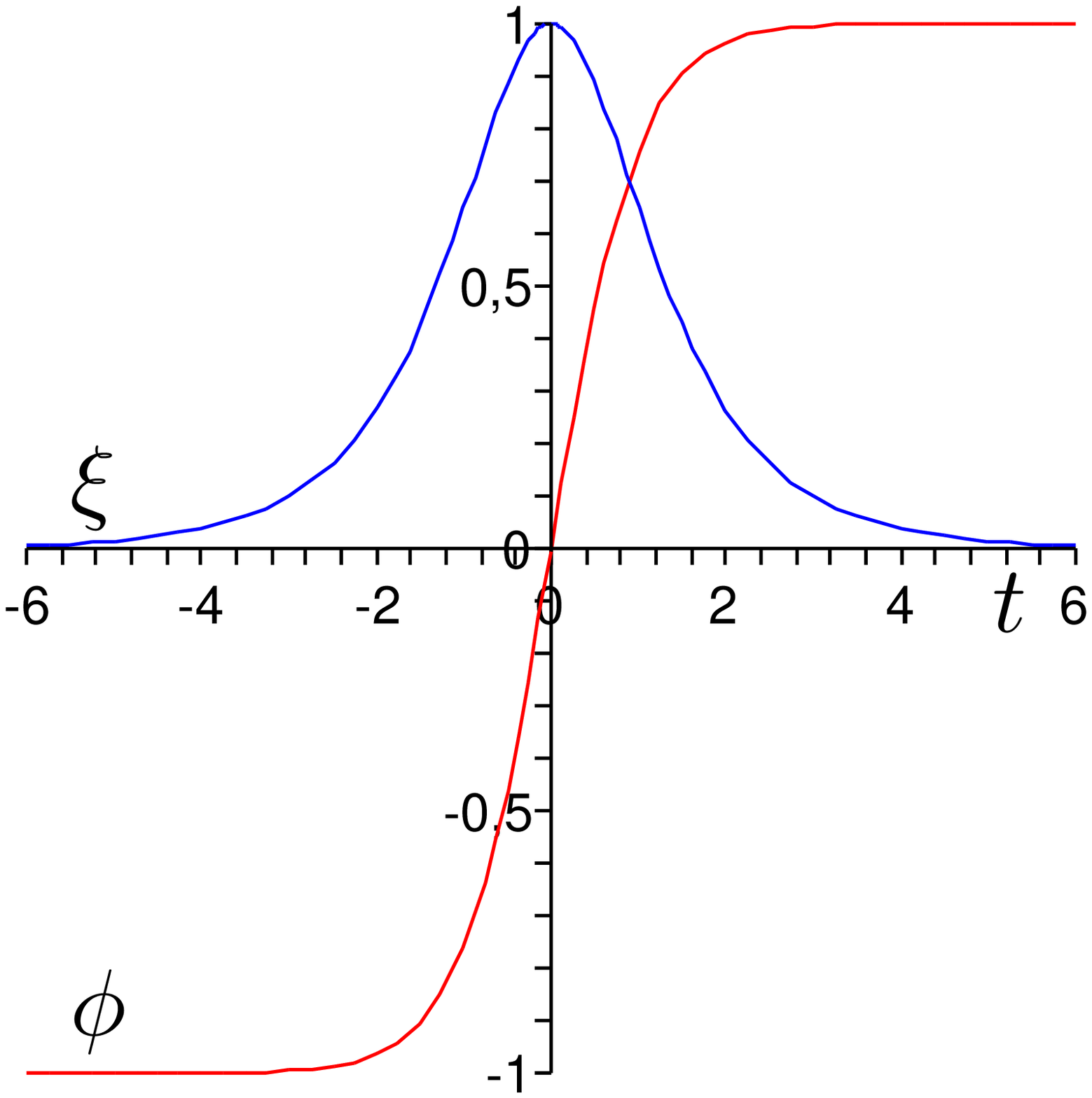} { \ \ }
\includegraphics[width=50mm] {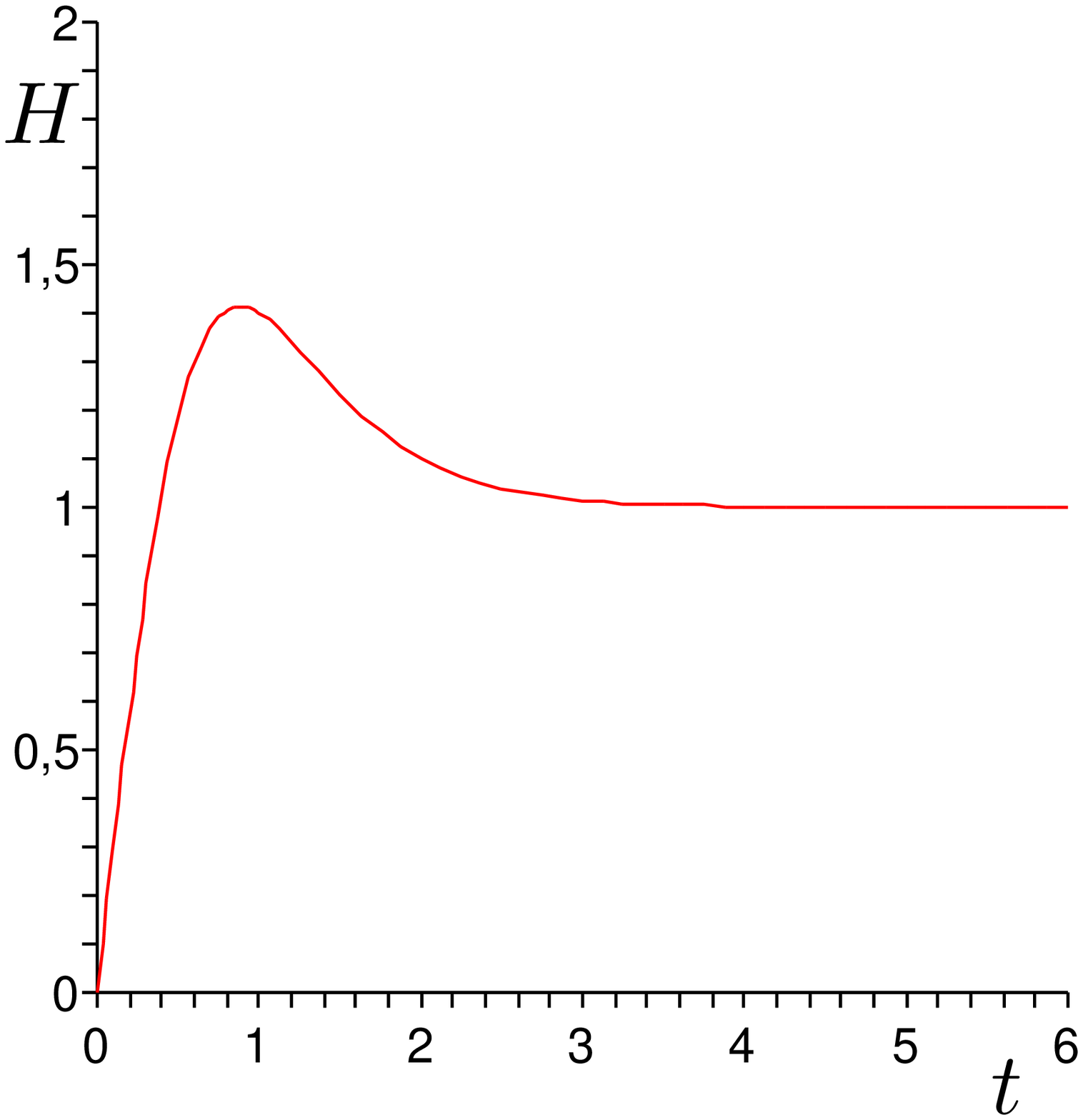}  { \ \ }
\includegraphics[width=50mm]{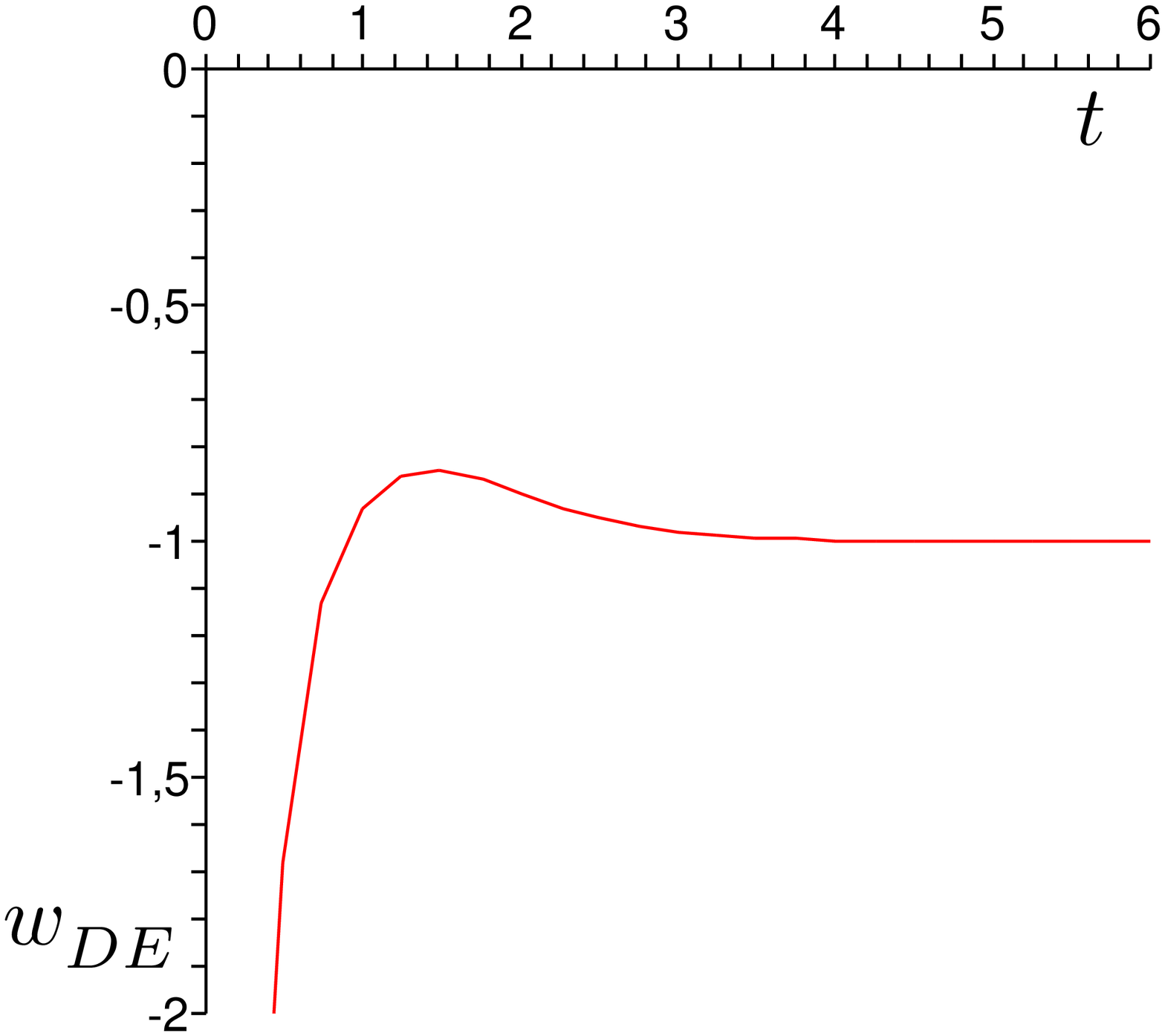}
\caption{The fields $\phi$ and $\xi$ (left), the Hubble parameter
$H$ (center) and the state parameter $w_{DE}$ (right) at $C=0$ and
$t_1=0$.} \label{H_th_ch}
\end{figure}

In the case $C=0$ we have solutions
\begin{equation}
\label{sol2t1}
    \phi_0(t)=A\tanh(\omega(t-t_1)) \qquad \mbox{and} \qquad
    \xi_0(t)={}\frac{A}{\cosh(\omega(t-t_1))}.
\end{equation}
At $t_1=0$ these solutions  coincide with solutions (\ref{sol2}).
The corresponding Hubble parameter
\begin{equation}
\label{Ht_th_ch}
 H_0=\frac{A^2\omega}{6m_p^2}\Bigl(3\tanh(\omega t)-2\tanh^3(\omega t)\Bigr)
\end{equation}
has a maximum at the point
$t_{max}={}-\ln\left(\sqrt{2}-1\right)/\omega\simeq 0.881/\omega$
and the quintessence large time behaviour. The solutions $\phi_0$
and $\xi_0$, the Hubble parameter $H_0$ and the state parameter
$w_{DE}^{\vphantom {27}}$ are presented  on \Fref{H_th_ch} (we put
$A=1$, $\omega=1$ and $m_p^2=1/6$).

For arbitrary $C$ the Hubble parameter is as follows:
\begin{equation}
\label{Hgen}
\begin{split}
H&=\frac{A^2\omega\left(e^{\omega(t_0-t)}+\left(C^2+4A^2\right)e^{\omega(t-t_0)}
\right)} {6m_p^2\left(e^{\omega(t_0-t)}
+2C+\left(C^2+4A^2\right)e^{\omega(t-t_0)}\right)^3}
\Bigl(e^{2\omega(t_0-t)}+6Ce^{\omega(t_0-t)}+{}\\&{}+10\left(C^2+4A^2\right)
+6C\left(C^2+4A^2\right)e^{\omega(t-t_0)}+\left(C^2+4A^2\right)^2e^{2\omega(t-t_0)}\Bigr).
\end{split}
\end{equation}

The straightforward calculations give that for all $C$, but
$C=\pm2A$, $\dot H(t)=0$ at four points
\begin{equation}
\label{tmax}
t_{m_k^{\vphantom{27}}}=t_0-\frac{1}{\omega}\ln\left(-\frac{4A^2+C^2\pm2A\sqrt{8A^2+2C^2}}{(C\pm
2A)(C^2+4A^2)}\right), \qquad k=1,\dots,4,
\end{equation}
where two signs "$\pm$" are independent. Note that if $C\neq \pm
2A$, then $\ddot H(t_{m_k^{\vphantom{27}}})\neq 0$. Therefore, the
Hubble parameter $H(t)$ has extrema at points
$t_{m_k^{\vphantom{27}}}$.

At $C>2A$ all four points $t_{m_k^{\vphantom{27}}}$ do not belong
to real axis.

If $C=2A$, then $\dot H(t)=0$ at two points, which do not belong
to real axis:
\begin{equation}
\tilde{t}_{m_1^{\vphantom{27}}}=t_0-\frac{1}{\omega}\ln(-2A)
\qquad\mbox{and} \qquad
\tilde{t}_{m_2^{\vphantom{27}}}=t_0-\frac{1}{\omega}\ln(-4A).
\end{equation}

So, at $C\geqslant 2A$ the Hubble parameter $H(t)$ is a
monotonically increasing function and its behavior is close to the
behavior of the Hubble parameter in one-field model~\cite{AKV}.

\begin{figure}[h]
\centering
\includegraphics[width=50mm]{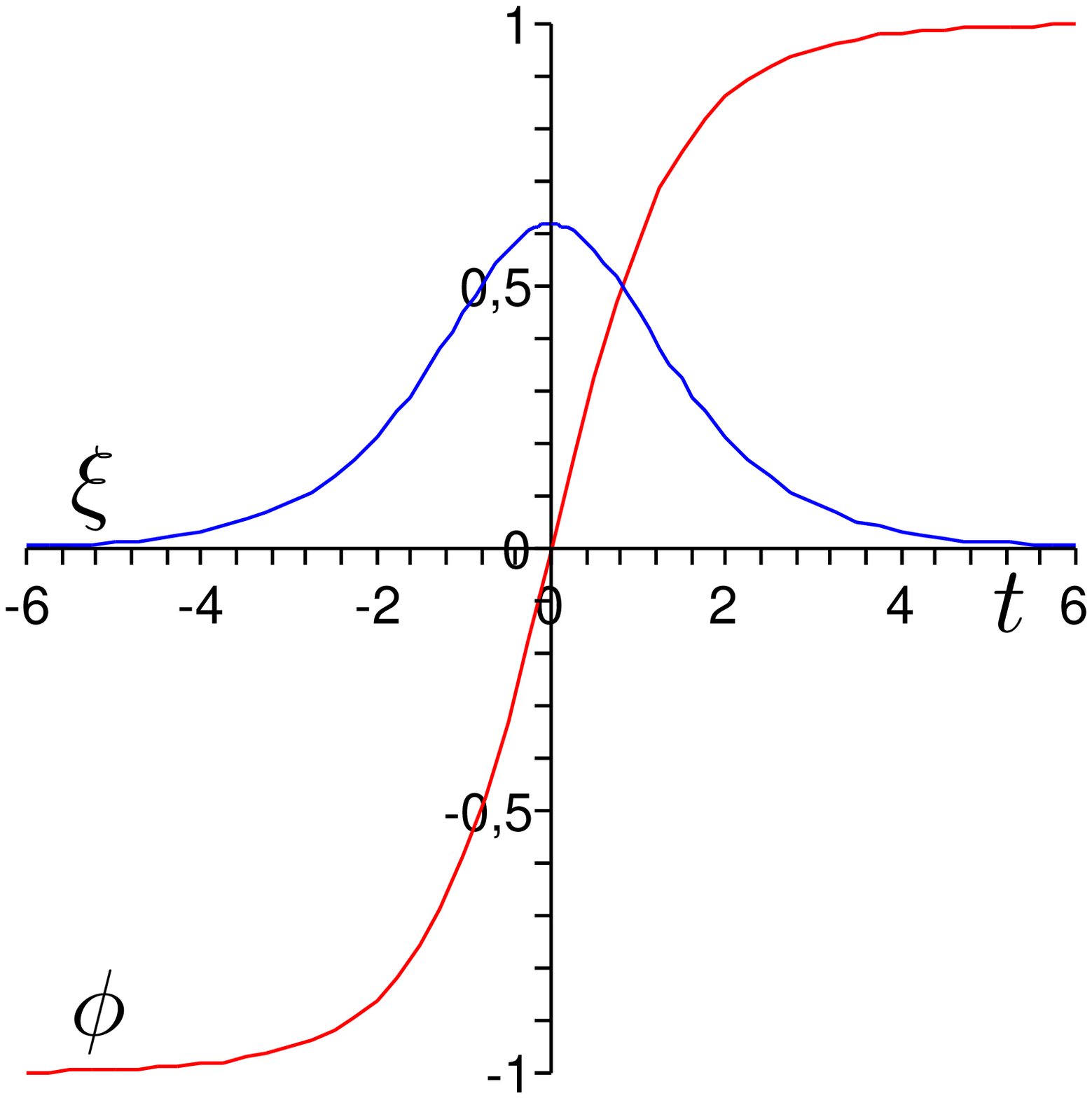} { \ \ }
\includegraphics[width=50mm]{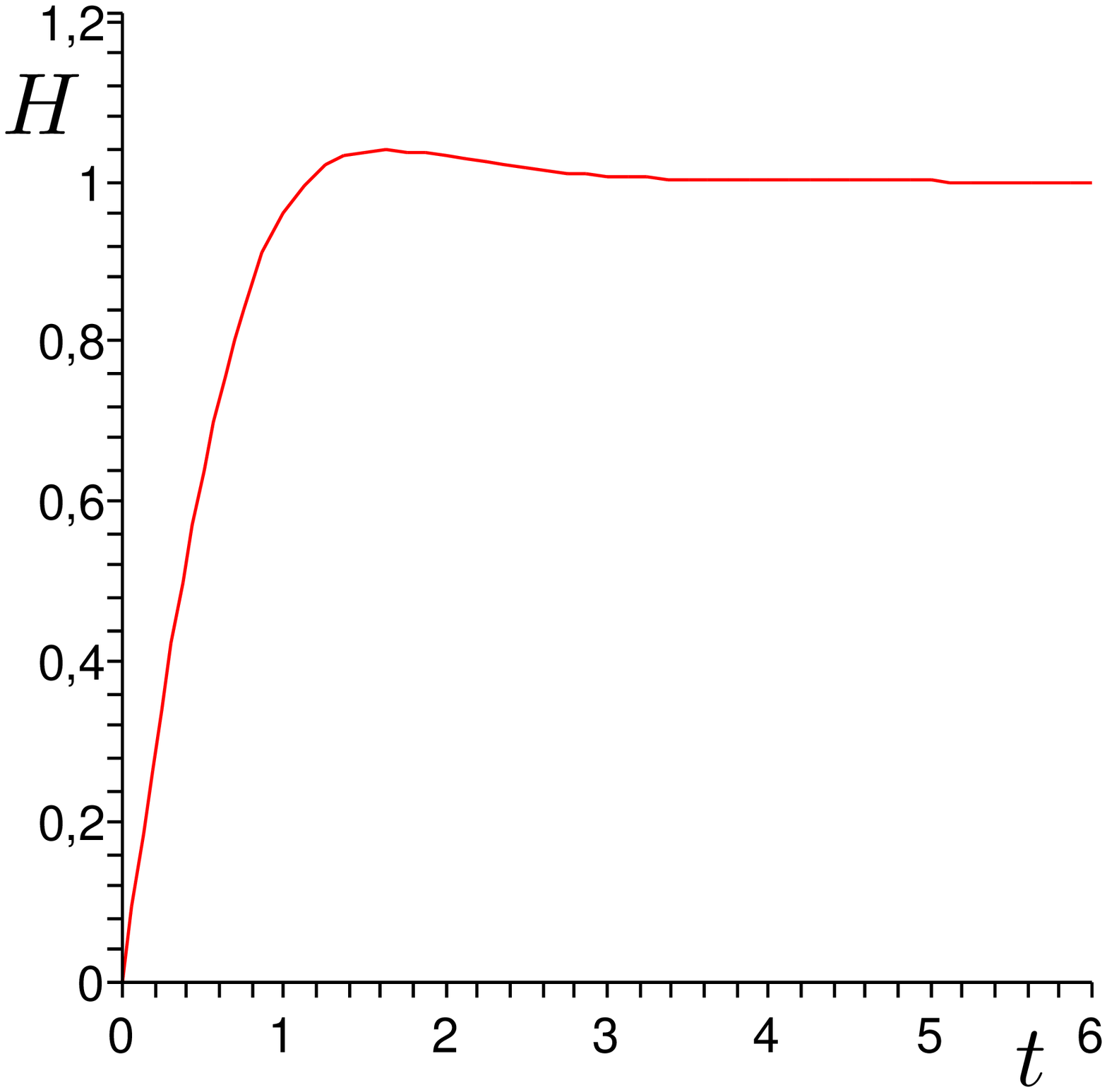} { \ \ }
\includegraphics[width=50mm]{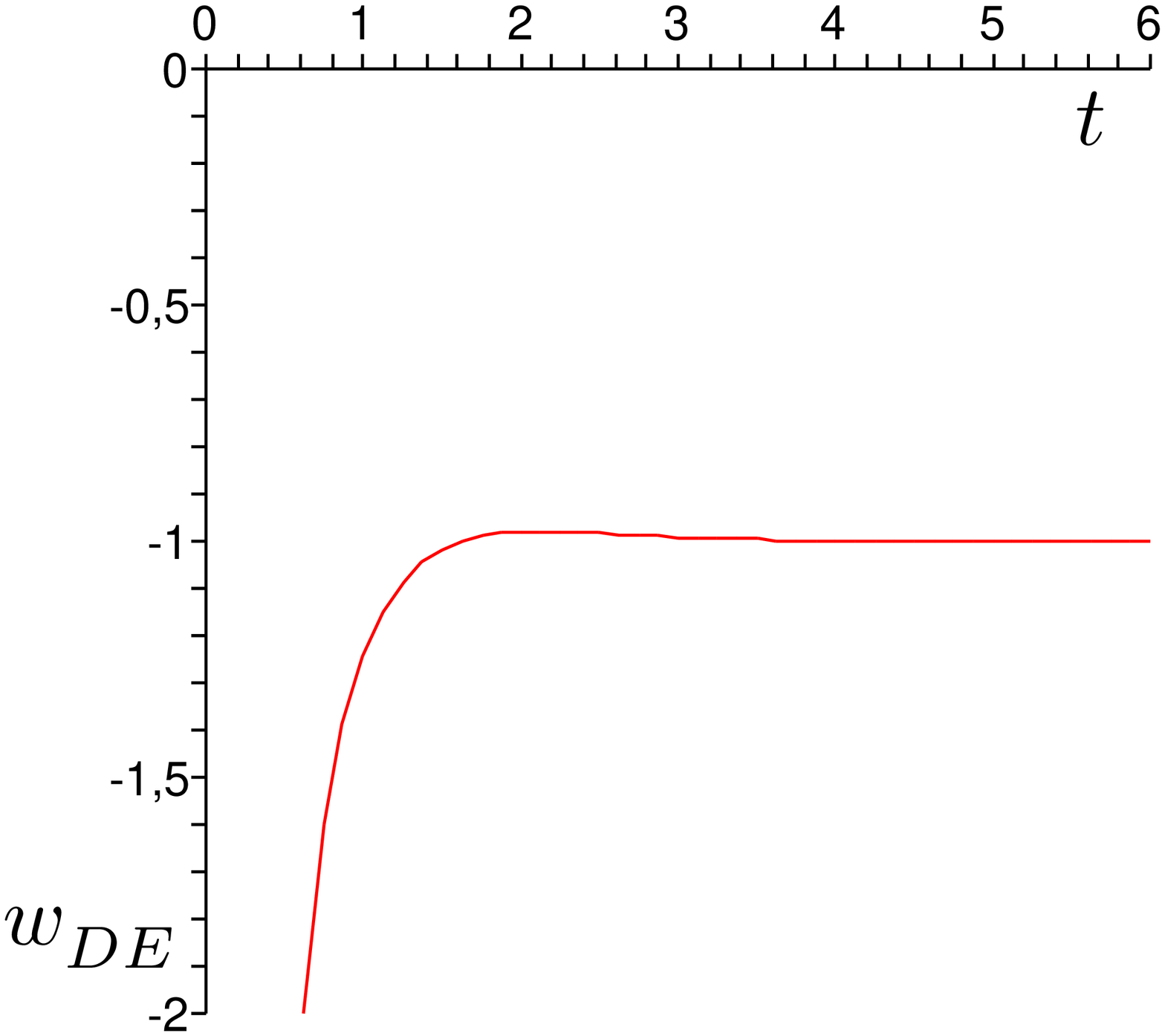}
\caption{The fields $\phi$ and $\xi$ (left), the Hubble parameter
$H$ (center) and the state parameter $w_{DE}$ (right) at $C=1$ and
$t_1=0$.} \label{Hpl}
\end{figure}

At $-2A<C<2A$ the function $H(t)$ has extrema at two points. If
$t_1=0$, the  $\phi(t)$ is an odd function, whereas $\xi(t)$ is an
even one. Therefore the corresponding Hubble parameter, calculated
by means of (\ref{W2}), is an odd function. It is easy to check
that on semi-axis $t>0$ the Hubble parameter $H(t)$ is positive
and, hence, has a maximum at $C<2A$ (see~\Fref{Hpl}). Thus, the
behavior of $H(t)$ in the case $-2A<C<2A$ looks like its behavior
at $C=0$.

If $C=-2A$, then $\dot H(t)=0$ at two points:
\begin{equation}
\tilde{t}_{m_3^{\vphantom{27}}}=t_0-\frac{1}{\omega}\ln(2A)
\qquad\mbox{and} \qquad
\tilde{t}_{m_4^{\vphantom{27}}}=t_0-\frac{1}{\omega}\ln(4A).
\end{equation}
At these points $\ddot H=\pm16A^2/m_p^2\neq 0$, hence, the Hubble
parameter behavior is close to $H(t)$ on Figures \ref{H_th_ch} and
\ref{Hpl}.

Let us consider the case $C<-2A$. All four points of extremum
(\ref{tmax}) are real. It means, that at $C<-2A$ we obtain a
qualitative new behavior of the Hubble parameter.

\begin{figure}[h]
\centering
\includegraphics[width=50mm]{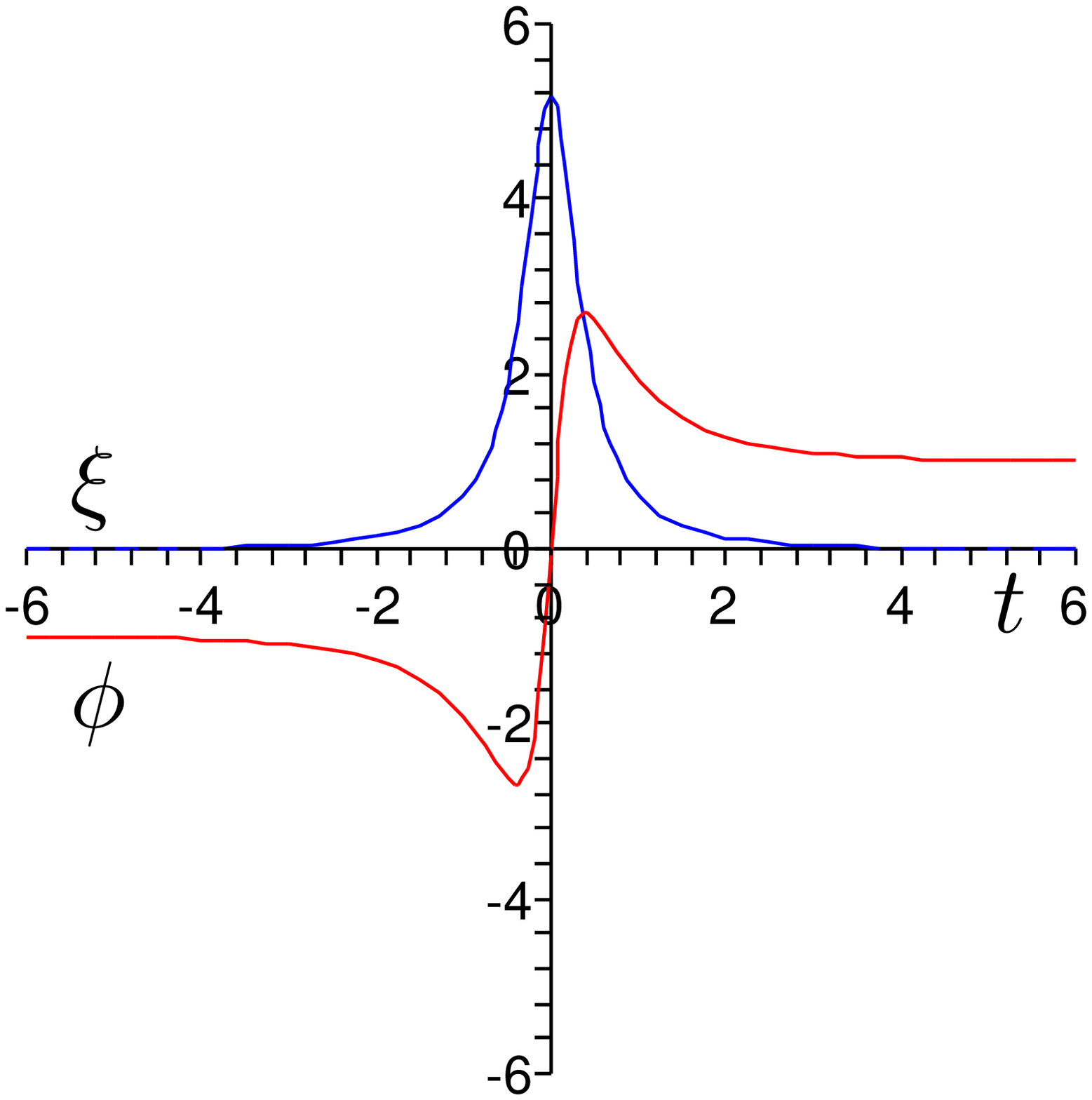} { \ \ }
\includegraphics[width=50mm]{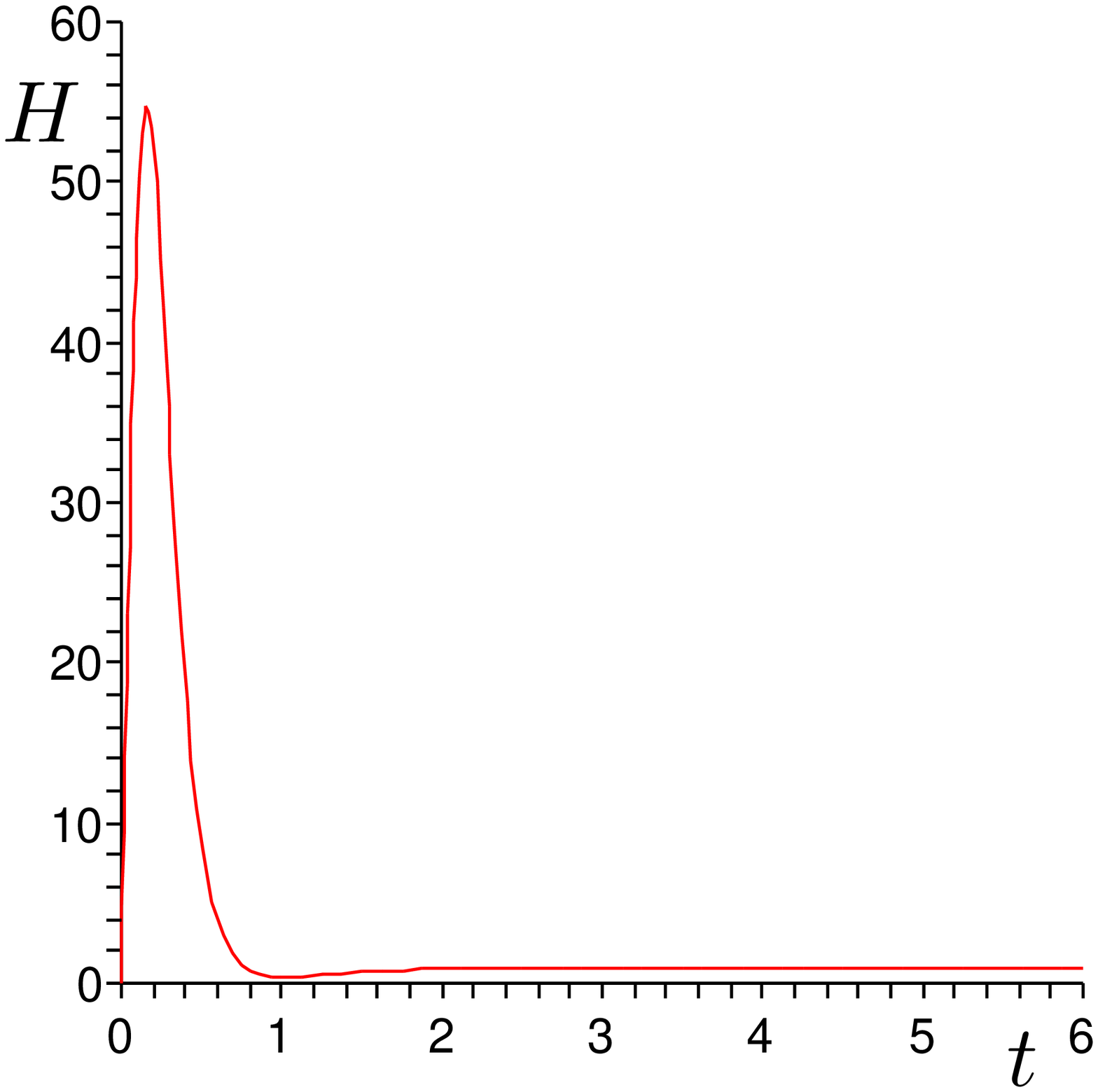} { \ \ }
\includegraphics[width=50mm]{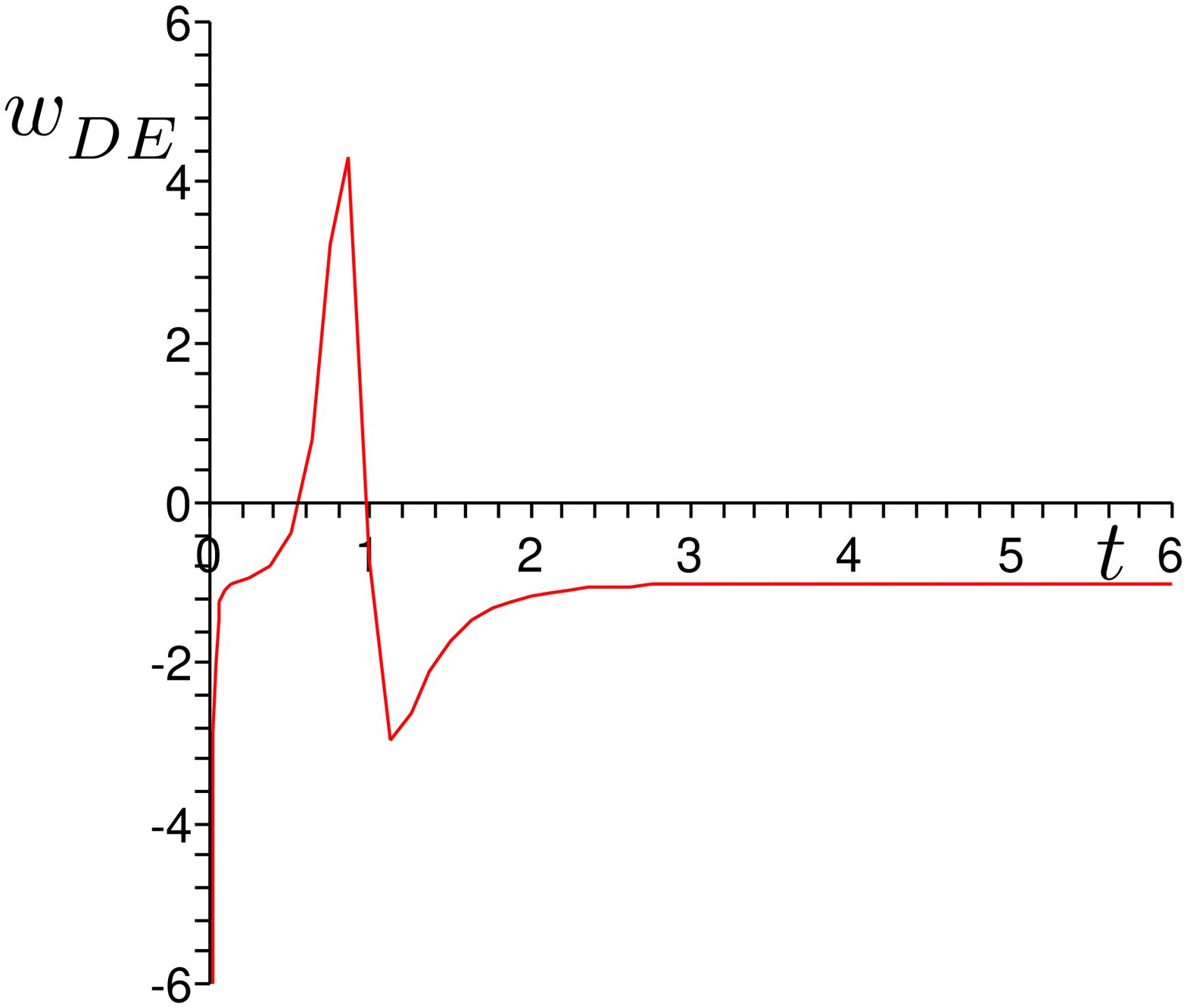}
\caption{The fields $\phi$ and $\xi$ (left), the Hubble parameter
$H$ (center) and the state parameter $w_{DE}$ (right) at $C=-5$
and $t_1=0$.} \label{Hphantom}
\end{figure}

If $t_1=0$, then, as it has been noted above, the Hubble parameter
is an odd function. The derivative of the Hubble parameter at zero
point is positive, hence, $H(t)$ has maximum at some point
$t_{m_1}>0$, a minimum at $t_{m_2}>t_{m_1}$ and is a monotonically
increasing function at $t>t_{m_2}$. Note that $w_{DE}^{\vphantom
{27}}<-1$ at $t>t_{m_2}$.  Thus we have found the exact solutions,
which correspond to the nonmonotonic function $H(t)$ with phantom
large time behaviour (see \Fref{Hphantom}).
Using the superpotential method we have obtained that the model
with the potential
\begin{equation}
\tilde{V}=\omega^2\left(\frac{1}{8}\left(1-\phi^2+\xi^2\right)^2
-\frac{1}{2A^2}\phi^2\xi^2+
\frac{3\phi^2}{4m_p^2}\left(\frac{A}{2}\left(1-\frac{\phi^2}{3A^2}\right)+
\frac{\xi^2}{2A}\right)^2\right) \label{V_Sergey}
\end{equation}
has two-parameter sets of exact solutions.  Note that the obtained
solutions have one and the same asymptotic conditions, whereas the
behaviour of the state parameter $w_{DE}^{\vphantom {27}}$ turn
out different. So, we can conclude that at large time both
quintessence ($w_{DE}^{\vphantom {27}}>-1$) and phantom
($w_{DE}^{\vphantom {27}}<-1$) behavior of $w_{DE}^{\vphantom
{27}}$ are possible to obtain from the SFT inspired effective
model with one and the same polynomial potential.

\section{Two-fields model with a polynomial potential and $w_{DE}$
crossing the cosmological barrier infinitely often}

From the Cubic Superstring Field Theory I.Ya.~Aref'eva and
A.S.~Koshelev obtained that the late time rolling of non-local
tachyon leads to a cosmic acceleration with a periodic crossing of
the cosmological constant barrier~\cite{AK}. At large time
approximation, when an open string tachyon
$\phi=1-\tilde{\delta\phi}$ and $|\tilde{\delta\phi}|\ll 1$, the
following Hubble parameter has been obtained:
\begin{equation}
\label{HAK}
    H=H_0+C_H^{\vphantom {27}}e^{-2rt}\sin(2\nu(t-t_0)),
\end{equation}
where $H_0$, $C_H^{\vphantom {27}}$, $r$, $\nu$ and $t_0$ are real
constants.

In~\cite{AK} the authors consider non-local model and the
corresponding Friedmann equations. In this paper we construct the
two-fields local model with the Hubble parameter (\ref{HAK}) in
the case $\nu=r$. For simplicity we put $C_H^{\vphantom
{27}}=1/(2m_p^2)$ and construct solutions, which do not depend on
$m_p^2$. In this case
\begin{equation}
\label{HAK2}
    \dot H=\frac{r}{m_p^2}e^{-2rt}
    \Bigl(2\sin(rt)^2-(\sin(rt)-\cos(rt))^2\Bigr).
\end{equation}
Using (\ref{eom2}) we can define the following explicit form of
solutions:
\begin{equation}
    \dot{\tilde{\delta\phi}}={}- 2\sqrt{r}e^{-rt}\sin(rt),
    \qquad \dot\xi=\sqrt{2r}e^{-rt}\Bigl(\sin(rt)-\cos(rt)\Bigr).
\end{equation}

It is easy to check that if
\begin{equation}
    \tilde{\delta\phi}= \frac{1}{\sqrt{r}}e^{-rt}\Big(\cos(rt)+\sin(rt)\Big),
    \qquad \xi={}-\frac{\sqrt{2}}{\sqrt{r}}e^{-rt}\sin(rt),
\end{equation}
then
\begin{equation}
\label{equphixiAK}
    \dot{\tilde{\delta\phi}}= \sqrt{2}r\xi, \qquad \dot\xi={}-\sqrt{2}r\tilde{\delta\phi}+2r\xi.
\end{equation}
Let construct the superpotential:
\begin{equation}
    \frac{\partial W}{\partial \tilde{\delta\phi}}
    =\frac1{2m_p^2}\sqrt{2}r\xi, \qquad
    \frac{\partial W}{\partial \xi}
    =\frac1{2m_p^2}\left(\sqrt{2}r\tilde{\delta\phi}-2r\xi\right),
\end{equation}
so
\begin{equation}
    W=\frac1{2m_p^2}\left(\sqrt{2}r\xi\tilde{\delta\phi}-r\xi^2\right)+H_0.
\end{equation}
The potential is (we put $H_0=0$)
\begin{equation}
\label{VAK}
    V=-r^2\left(\xi^2+\tilde{\delta\phi}^2-2\sqrt{2}\xi\tilde{\delta\phi}\right)+\frac{3r^2}{4m_p^2}
    \left(\sqrt{2}\xi\tilde{\delta\phi}-\xi^2\right)^2.
\end{equation}

Thus we obtain the explicit solutions and the fourth degree
polynomial potential, which corresponds to the Hubble parameter
from the SFT inspired model with high derivatives.

Note that the standard method to construct models with scalar
fields for the given behaviour of the Hubble parameter is the
method, which uses $V(\phi,\xi)$ as a function of
time~\cite{Andrianov}. If we know the Hubble parameter $H(t)$,
then, using (\ref{Vt}), we obtain $V(t)$ and after that we can
attempt to find the functions $\phi(t)$ and $\xi(t)$ and the
potential $V(\phi,\xi)$. Such method is very effective if at least
one of derivatives either $V_\phi$ or $V_\xi$ is  such a function
$F(V)$ that a form of $F$ does not depend on $\phi$ and $\xi$. For
example, if
\begin{equation}
\label{Vexp}
    V(\phi,\xi)=V_1(\phi)e^{\alpha\xi},
\end{equation}
where $\alpha$ is a constant, then
\begin{equation}
    \frac{\partial V(\phi(t),\xi(t))}{\partial\xi}=\alpha V_1(\phi(t))
    e^{\alpha\xi(t)}=\alpha V(t)
\end{equation}
and (\ref{eom4}) is a linear differential equation in $\xi$
\begin{equation}
\label{eom4Andr}
    \ddot{\xi}+3H(t)\dot{\xi}+\alpha\left(3H^2(t)+\dot{H}(t)\right)=0.
\end{equation}
This equation allows to find $\xi(t)$ if the Hubble parameter
$H(t)$ is known~\cite{Andrianov}.  The superpotential method is
not so effective to seek potential in the form (\ref{Vexp}) for
the given Hubble parameter $H(t)$.  On the other hand if the
required form of the potential is a polynomial, the superpotential
method is not less effective and maybe even more easy to use than
the above-mentioned method.

\section{Conclusions}
In this paper we have investigated the dynamics of two component
DE models, with one phantom field and one usual field. The main
motivation for us is a model of the Universe as a slowly decaying
D3-brane, whose dynamics is described by a tachyon
field~\cite{Arefeva}. To take into account the back reaction of
gravity we add a scalar field with an usual kinetic term.

We construct a cosmological model with the SFT inspired polynomial
potential $V(\phi,\xi)$ and find two-parameter set of exact
solutions. This set can be separated into two subset such that one
subset corresponds to the quintessence large time behaviour,
another subset corresponds to the phantom one. Note that both
subsets have solutions, which satisfy one and the same asymptotic
conditions and the additional condition $\phi(0)=0$.

We also construct two-fields model with the fourth degree
polynomial potential, which corresponds to the Hubble parameter,
obtained in the SFT framework~\cite{AK}. In this model the state
parameter $w_{DE}$ crosses the cosmological constant barrier
infinitely often.

In this paper we actively use the superpotential method and show
that there are new ways to use this method in the case of two
fields. We can not only  construct potential for the given
solutions, but also find new solutions. In particular
superpotential method allows to generalize a one-parameter set of
solutions up to two-parameter set. The superpotential method
allows to separate the initial system of motion
equations~(\ref{eom1})--(\ref{eom4}) into two parts. One part is
the equation on superpotential~(\ref{deWolfe_potential}), which in
general case is not integrable, but for many polynomial potentials
has special solutions. Substituting these solutions into the
second part
(system~(\ref{deWolfe_method1})--(\ref{deWolfe_method2})) we
obtain a system of ordinary differential equations, which is
usually integrable at least in quadratures. Note that the systems
of the type (\ref{deWolfe_method1})--(\ref{deWolfe_method2}) are
actively investigated both in mechanics and in supersymmetry
theories with BPS states. So, the superpotential method allows to
stand out from the system of the Friedmann equations a subsystem,
which can be integrable, even in the case of a nonintegrable
initial system. On the other hand this method allows to make such
a fine tuning of parameters of the considering gravitational
models, for example, a choose of coefficients of the potential,
that the explicit solutions exist.

For solutions with $H(0)=0$ we obtain that $\dot H(0)>0$. Similar
solutions are known as bounce ones (see, for
example,~\cite{Cai07}). Note, that bounce solutions has been
obtained in the SFT inspired
higher-derivative models~\cite{Biswas,AJV}.

\section*{Acknowledgments}
Author is grateful to A.A.~Andrianov, I.Ya.~Aref'eva and
A.Yu.~Kamenshchik  for useful discussions. This research is
supported in part by RFBR grant 05-01-00758 and by Russian
Fede\-ration President's grant NSh--8122.2006.2.


\begin{thebibliography}{99}
\bibitem{Riess1}  A. Riess {\it et al.}
 [Supernova Search Team Collaboration],
{\it Observational Evidence from Supernovae for an Accelerating
Universe  and a Cosmological Constant}, \AJ {\bf 116} (1998)
1009--1038; astro-ph/9805201.


\bibitem{Spergel}
 D.N. Spergel {\it et al.}  [WMAP Collaboration],
 {\it First Year Wilkinson Microwave Anisotropy Probe (WMAP) Observations:
 Determination of  Cosmological Parameters},  \APJ Suppl. {\bf 148} (2003)
 175--194; astro-ph/0302209,

 D.N. Spergel {\it et al.}  [WMAP Collaboration],
  \textit{Wilkinson Microwave Anisotropy Probe (WMAP) Three
 Year Results: Implications for  Cosmology},
 Astrophys. J. Suppl. Ser. \textbf{170} (2007) 377--408; astro-ph/0603449.

\bibitem{Tegmark} Tegmark {\it et al.}  [SDSS Collaboration],
\textit{Cosmological parameters from SDSS and WMAP,}
 Phys. Rev. \textbf{D69} (2004) 103501; astro-ph/0310723.

\bibitem{Astier}  P. Astier {\it et al.}  [SNLS Collaboration],
The Supernova Legacy Survey: Measurement of $\Omega_M$,
$\Omega_\Lambda$ and $w$ from the First Year Data Set,
\textit{Astron. Astrophys.} \textbf{447} (2006) 31--48;
astro-ph/0510447.

 \bibitem{Copeland} E.J. Copeland, M. Sami, Sh. Tsujikawa,
 \textit{Dynamics of dark
energy}, \textit{Int. J. Mod. Phys.}   {\bf D15} (2006)
1753--1936, hep-th/0603057

\bibitem{Gong} Y. Gong, A. Wang, \textit{Reconstruction of the deceleration parameter and
the equation of state of dark energy}, astro-ph/0612196.

\bibitem{ASS} U. Alam, V. Sahni,
A.A. Starobinsky, \textit{Exploring the Properties of Dark Energy
Using Type Ia Supernovae and Other Datasets}, JCAP \textbf{0702}
(2007) 011, astro-ph/0612381

\bibitem{VaStar} V. Sahni, A.A. Starobinsky, \textit{Reconstructing
Dark Energy}, Int. J. Mod. Phys. \textbf{D15} (2006) 2105--2132,
astro-ph/0610026.

 \bibitem{AKV} I.Ya.~Aref'eva,
A.S.~Koshelev, S.Yu.~Vernov, {\it Exact Solvitions in a String
Cosmological Model}, Theor. Math. Phys. \textbf{148} (2006)
895--909 [Teor. Mat. Phys. \textbf{148} (2006) 23--41];
astro-ph/0412619.

\bibitem{VarunStar}  U. Alam, V. Sahni, T.D. Saina A.A.
Starobinsky, \textit{Is there Supernova Evidence for Dark Energy
Metamorphosis?},
 \textit{Mon. Not. Roy. Astron. Soc.} \textbf{354} (2004) 275;
astro-ph/0311364

\bibitem{Vikman}
A. Vikman, {\it Can dark energy evolve to the Phantom?}, Phys.
Rev.  D {\bf 71} (2005) 023515; astro-ph/0407107.

\bibitem{across-1}
Sh. Nojiri, S.D. Odintsov, \textit{Modified gravity and its
reconstruction from the universe expansion history},
hep-th/0611071,

Sh. Nojiri, S.D. Odintsov, {\it The new form of the equation of
state for dark energy fluid and accelerating
 universe}, Phys. Lett. \textbf{B639} (2006) 144--150; hep-th/0606025,


Sh. Nojiri, S.D. Odintsov, {\it Unifying phantom inflation with
late-time acceleration: scalar phantom-non-phantom transition
model and generalized holographic dark energy}, Gen. Rel. Grav.
{\bf 38} (2006) 1285--1304; hep-th/0506212,

Sh. Nojiri, S.D. Odintsov, {\it Inhomogeneous equation of state of
the universe: phantom era, future singularity and crossing the
phantom barrier}, Phys. Rev. {\bf D72} (2005) 023003;
hep-th/0505215,

Sh. Nojiri, S.D. Odintsov, Sh. Tsujikawa, \textit{Properties of
singularities in (phantom) dark energy universe}, Phys. Rev.
\textbf{D71} (2005) 063004; hep-th/0501025,

Sh. Tsujikawa 2005, \textit{Reconstruction of general scalar-field
dark energy models}, Phys. Rev.  \textbf{D72} (2005) 083512;
astro-ph/0508542,

S. Nesseris, L. Perivolaropoulos, \textit{Crossing the Phantom
Divide: Theoretical Implications and Observational Status}, JCAP
\textbf{0701} (2007) 018; astro-ph/0610092,

R. Gannouji, D. Polarski, A. Ranquet, A.A.
 Starobinsky, \textit{Scalar-Tensor Models of Normal and Phantom Dark
 Energy}, J. Cosmol. Astropart. Phys.  \textbf{0609} (2006) 016;
 astro-ph/0606287,

V.A. Rubakov, \textit{Phantom without UV pathology}, Theor. Math.
Phys. \textbf{149} (2006) 1651--1664 (Teor. Mat. Fiz. \textbf{149}
(2006) 409--426), hep-th/0604153,

H. Mohseni Sadjadi, M. Alimohammadi, \textit{Cosmological
coincidence problem in interacting dark energy models}, Phys. Rev.
\textbf{D74} (2006) 103007; gr-qc/0610080,

M. Alimohammadi, H. Mohseni Sadjadi, \textit{The $w ={} -1$
crossing of the quintom model with arbitrary potential},
gr-qc/0608016,

H. Mohseni Sadjadi, M. Alimohammadi, \textit{ Transition from
quintessence to phantom phase in quintom model}, Phys. Rev.
\textbf{D74} (2006) 043506;  gr-qc/0605143,

 P.S. Apostolopoulos, N. Tetradis,
 \textit{Late acceleration and $w=-1$ crossing in induced gravity},
  Phys. Rev.  \textbf{D74} (2006)
 064021; hep-th/0604014,

 V. Sahni,   Yu. Shtanov, \textit{Brane World Models of Dark
Energy}, \textit{JCAP} \textbf{11} (2003) 014,
 astro-ph/0202346,

Wen Zhao, Yang Zhang, \textit{The Quintom Models with State
Equation Crossing $-1$}, Phys. Rev. \textbf{D73} (2006) 123509;
astro-ph/0604460,

Bo Feng, Xiulian Wang, Xinmin Zhang,  \textit{Dark Energy
Constraints from the Cosmic Age and Supernova}, Phys. Lett.
\textbf{B607} (2005) 35--41; astro-ph/0404224,

Zong-Kuan Guo, Yun-Song Piao, Xinmin Zhang, Yuan-Zhong Zhang, {\it
Cosmological Evolution of a Quintom Model of Dark Energy}, Phys.
Lett. \textbf{B608} (2005) 177--182; astro-ph/0410654,

 Xiao-Fei Zhang, Hong Li, Yun-Song Piao,
{\it Two-field models of dark energy with equation of state across
$-1$}, Mod. Phys. Lett. \textbf{A21} (2006) 231--242;
 astro-ph/0501652,

 Ming-zhe Li, Bo Feng, Xin-min Zhang, {\it A single scalar field model
of dark energy with equation of state crossing $-1$},
hep-ph/0503268,

 Hrv. Stefancic, {\it Dark energy transition between quintessence and
phantom regimes --- an equation of state analysis}, Phys. Rev.
\textbf{D71} (2005) 124036;  astro-ph/0504518,

 Rong-Gen Cai, Hong-Sheng Zhang, Anzhong Wang, {\it Crossing $w=-1$ in
Gauss-Bonnet Brane World with Induced Gravity}, Commun. Theor.
Phys. \textbf{44} (2005) 948--954; hep-th/0505186,

Yi-Fu Cai, Hong Li, Yun-Song Piao, Xinmin Zhang,  \textit{Cosmic
Duality in Quintom Universe},  gr-qc/0609039,

Xin Zhang, \textit{Dynamical vacuum energy, holographic quintom,
and the reconstruction of scalar-field dark energy},  Phys. Rev.
\textbf{D74} (2006) 103505;  astro-ph/0609699.

Hongsheng Zhang, Zong-Hong Zhu, \textit{Crossing $w=-1$ by a
single scalar on a Dvali-Gabadadze-Porrati brane},
 Phys.Rev. \textbf{D75} (2007) 023510, astro-ph/0611834

\bibitem{Andrianov} A.A. Andrianov, F. Cannata, A.Yu. Kamenshchik, \textit{Complex
Lagrangians and phantom cosmology}, J. Phys. \textbf{A39} (2006)
9975--9982; gr-qc/0604126.

\bibitem{AK} I.Ya.~Aref'eva, A.S.~Koshelev, {\it Cosmic acceleration
and crossing of $w={}-1$ barrier in non-local Cubic
  Superstring Field Theory model}, JHEP \textbf{0702} (2007) 041;
 hep-th/0605085

\bibitem{wless-1}

 V.K. Onemli,  R.P. Woodard,
 Super-Acceleration from Massless, Minimally Coupled $\phi^4$,
Class. Quantum Grav. \textbf{19} (2002) 4607; gr-qc/0204065,

S.M. Carroll, M. Hoffman, M. Trodden, {\it  Can the dark energy
equation-of-state parameter $w$ be less than $-1$?}, Phys. Rev.
\textbf{D68} (2003) 023509; astro-ph/0301273.

St.D.H. Hsu, A. Jenkins, M.B. Wise, \textit{Gradient instability
for $w < -1$}, Phys. Lett. \textbf{B597} (2004) 270--274;
astro-ph/0406043,

R.V. Buniy, St.D.H. Hsu, B.M. Murray, \textit{The null energy
condition and instability},  Phys. Rev. \textbf{D74} (2006)
063518, hep-th/0606091

B. McInnes, \textit{Phantom Divide in String Gas Cosmology},
 Nucl. Phys. \textbf{B718} (2005) 55--82; hep-th/0502209

V.~Gorini, A.Yu. Kamenshchik, U. Moschella, V. Pasquier, A.A.
Starobinsky, \textit{Stability properties of some perfect fluid
cosmological models}, Phys. Rev. \textbf{D72} (2005) 103518;
astro-ph/0504576.

E.O. Kahya, V.K. Onemli, \textit{Quantum Stability of a $w < - 1$
Phase of Cosmic Acceleration}, Phys. Rev. \textbf{D76} (2007)
043512, gr-qc/0612026.

\bibitem{AreVo}
I.Ya.~Aref'eva, I.V. Volovich, \textit{On the Null Energy
Condition and Cosmology}, hep-th/0612098.

\bibitem{Arefeva}
 I.Ya. Aref'eva,
 {\it Nonlocal String Tachyon as a Model for Cosmological Dark Energy},
 AIP Conf. Proc. \textbf{826} (2006) 301--311, astro-ph/0410443.

\bibitem{SFT-review}
K.~Ohmori, \textit{A Review on Tachyon Condensation in Open String
Field Theories}, hep-th/0102085;

I.Ya.~Aref'eva, D.M.~Belov, A.A.~Giryavets, A.S.~Koshelev,
P.B.~Medvedev, \textit{Noncommutative Field Theories and
(Super)String Field Theories}, hep-th/0111208;

W. Taylor, {\it Lectures on D-branes, tachyon condensation and
string field theory}, hep-th/0301094.

\bibitem{AKV3} I.Ya. Aref'eva, A.S. Koshelev, S.Yu.
Vernov, \textit{Crossing of the $w=-1$ Barrier by D3-brane Dark
Energy Model}, Phys.  Rev. \textbf{D72} (2005) 064017;
astro-ph/0507067.


\bibitem{AJ} I.Ya.~Aref'eva, L.V.~Joukovskaya, {\it Time Lumps in
 Nonlocal Stringy Models
and Cosmological Applications},  J. High Energy Phys. {\bf 0510}
(2005) 087; hep-th/0504200.

\bibitem{LY} L.V.~Joukovskaya, Ya.I.~Volovich, {\it
Energy Flow from Open to Closed Strings in a Toy Model of Rolling
Tachyon}, math-ph/0308034.

\bibitem{AKV2} I.Ya.~Aref'eva,
A.S.~Koshelev, S.Yu.~Vernov, \textit{Stringy Dark Energy Model
with Cold Dark Matter}, Phys. Lett. \textbf{B628} (2005) 1--10;
astro-ph/0505605.

\bibitem{AKVBulg}I.Ya.~Aref'eva,
A.S.~Koshelev, S.Yu.~Vernov, \textit{Exact Solutions in $w<-1$ SFT
Inspired  Cosmological Models}, Bulgarian J. of Phys., \textbf{33,
Suppl. 1a} (2006)  360--367.

\bibitem{ABKM1} I.Ya.~Arefeva, D.M.~Belov, A.S.~Koshelev,
P.B.~Medvedev, \textit{Tachyon condensation in cubic superstring
field theory}, Nucl. Phys  {\bf B638} (2002) 3--20;
hep-th/0011117.

\bibitem{ABKM2} I.Ya.~Arefeva, D.M.~Belov, A.S.~Koshelev,
P.B.~Medvedev, \textit{Gauge invariance and tachyon condensation
in cubic superstring field  theory}, Nucl. Phys  {\bf B638} (2002)
21--40; hep-th/0107197.

\bibitem{Witten-SFT}  E.~Witten,
\textit{Noncommutative geometry and string field theory}, Nucl.
Phys. \textbf{B268} (1986) 253--294;

E.~Witten, \textit{Interacting field theory of open superstrings},
Nucl.Phys.~\textbf{B276} (1986) 291--324.

\bibitem{AMZ-PTY}  I.Ya.~Aref'eva, P.B.~Medvedev, A.P.~Zubarev,
\textit{Background formalism for superstring field theory}, Phys.
Lett.~\textbf{B240} (1990) 356--362;


C.R.~Preitschopf, C.B.~Thorn, S.A.~Yost, \textit{Superstring Field
Theory}, Nucl. Phys. \textbf{B337} (1990) 363--433;


I.Ya.~Aref'eva,  P.B.~Medvedev, A.P.~Zubarev, \textit{New
representation for string field solves the consistency problem for
open superstring field}, Nucl. Phys.~\textbf{B341} (1990)
464--498.

\bibitem{BSZ} N. Berkovits, A. Sen, B. Zwiebach,
\textit{Tachyon Condensation in Superstring Field Theory}, Nucl.
Phys. \textbf{B587} (2000) 147--178,  hep-th/0002211.

\bibitem{AJK} I.Ya.~Aref'eva, L.V.~Joukovskaya, A.S.~Koshelev,
\textit{Time Evolution in Superstring Field Theory on non-BPS
brane. Rolling Tachyon and Energy-Momentum Conservation},
hep-th/0301137.

I.Ya. Aref'eva, {\it Rolling tachyon in NS string field theory},
Fortschr. Phys. \textbf{51} (2003) 652--657.

\bibitem{Oh}  K. Ohmori, {\it Toward open closed string theoretical
    description of rolling tachyon},
    Phys. Rev. \textbf{D69} (2004) 026008; hep-th/0306096.

\bibitem{BZ} B. Zwiebach, {\it Oriented open-closed string theory
revisited}, Annals Phys. \textbf{267} (1998)  193--248;
hep-th/9705241

\bibitem{DeWolfe} O. DeWolfe, D.Z. Freedman, S.S. Gubser, A. Karch,
{\it Modeling the fifth dimension with scalars and gravity},
 Phys. Rev. {\bf D62} (2000) 046008, hep-th/9909134.

\bibitem{Muslimov} A.G. Muslimov, \textit{On the Scalar Field Dynamics
in a Spatially Flat Friedman Universe}, Class. Quant. Grav.
\textbf{7} (1990) 231--237.

\bibitem{Salopek}
D.S. Salopek, J.R. Bond, \textit{Nonlinear evolution of
long-wavelength metric fluctuations in inflationary models}, Phys.
Rev. \textbf{D42} (1990) 3936--3962.

\bibitem{Liddle} A.R. Liddle, D.H. Lyth, \textit{Cosmological Inflation
and Large-scale Structure}, Cambridge, NY, 2000.

\bibitem{Bazeia95} D. Bazeia, M.J. dos Santos, R.F. Ribeiro,
\textit{Solitons in systems of coupled scalar fields}, Phys. Lett.
\textbf{A208} (1995) 84--88; hep-th/0311265.

\bibitem{Bazeia99}  D. Bazeia, F.A. Brito,
\textit{Bags, junctions, and networks of BPS and non-BPS defects},
Phys. Rev. \textbf{D61} (2000) 105019; hep-th/9912015.

\bibitem{Brandhuber} A. Brandhuber, K. Sfetsos, \textit{Non-standart
compatifications with mass gaps ans Newton's law}, J. High Energy
Phys. \textbf{9910} (1999) 013; hep-th/9908116.

\bibitem{Bogomolnyi} E.B. Bogomol'nyi,
\textit{Stability Of Classical Solutions}, Sov. J. Nucl. Phys.
\textbf{24} (1976) 449, Yad. Fiz. \textbf{24} (1976) 861--870.

\bibitem{BazeiaDE}
D. Bazeia, C.B. Gomes, L. Losano, R. Menezes, \textit{First-order
formalism and dark energy}, Phys. Lett. \textbf{B633} (2006)
415--419; astro-ph/0512197;

D. Bazeia, L. Losano, J.J. Rodrigues, \textit{First-order
formalism for scalar field in cosmology}, hep-th/0610028

\bibitem{BazeiaCDM}
D. Bazeia, L. Losano, R. Rosenfeld, \textit{First-order formalism
for dust}, astro-ph/0611770


\bibitem{MMVS} A.S. Mikhailov, Yu.S. Mikhailov, M.N. Smolyakov, I.P. Volobuev,
\textit{Constructing stabilized brane world models in
five-dimensional Brans-Dicke theory}, Class. Quantum Grav.
\textbf{24} (2007) 231--242; hep-th/0602143.

\bibitem{Cai07} Yi-Fu Cai, Taotao Qiu, Yun-Song Piao,
Mingzhe Li, Xinmin Zhang, \textit{Bouncing universe with quintom
matter}, arXiv:0704.1090.

\bibitem{Biswas} T. Biswas, A. Mazumdar, W. Siegel,
\textit{Bouncing Universes in String-inspired Gravity},
    JCAP \textbf{0603} (2006) 009; hep-th/0508194

\bibitem{AJV} I.Ya.~Aref'eva, L.V.~Joukovskaya, S.Yu.~Vernov,
\textit{Bouncing and Accelerating Solutions in Nonlocal Stringy
Models,} J. High Energy Phys. \textbf{0707} (2007) 087;
hep-th/0701184.

\end{thebibliography}
\end{document}